**Title:** A Comparative Study of Neutron Irradiation for Genetic Mutations: Spallation, Reactor, and Compact Neutron Source

**Authors:** May Sweet (1 and 2), Kenji Mishima (3 and 2), Masahide Harada (4), Keisuke Kurita (5), Hiroshi Iikura (5), Seiji Tasaki (6), Norio Kikuchi (1 and 2)

**Affiliations:**
(1) R & D Department, Quantum Flowers & Foods Co.Ltd, QFF Tokai Lab, Quantum Beam Research Center 2 Room 101, AYA'S LABORATORY 162-1 Shirakata, Tokai-mura, Naka-gun, Ibaraki, 319-1106, Japan.
(2) Institute of Materials Structure Science, High Energy Accelerator Research Organization, KEK, 203-1 Shirakata, Tokai-mura, Naka-gun, Ibaraki, 319-1106, Japan.
(3) Institute for the Origin of Particles and the Universe (KMI), International Research Center for Flavor Physics, Nagoya University, Furocho, Chikusa Ward, Nagoya City, Aichi, 464-8602, Japan.
(4) J-PARC Center, Japan Atomic Energy Agency (JAEA), 2-4 Shirakata, Tokai, Naka, Ibaraki, 319-1195, Japan.
(5) Materials Sciences Research Center, Japan Atomic Energy Agency (JAEA), 2-4 Shirakata, Tokai, Naka, Ibaraki, 319-1195, Japan.
(6) Department of Nuclear Engineering, Graduate School of Engineering, Kyoto University, Kyoto University Katsura, Nishikyo-ku, Kyoto, 615-8530, Japan.

**Corresponding Author:**
May Sweet
Researcher, R & D department
Quantum Flowers & Foods Co.Ltd,
QFF Tokai Lab, Quantum Beam Research Center 2 Room 101,
AYA'S LABORATORY 162-1 Shirakata, Tokai-mura, Naka-gun,
Ibaraki, 319-1106, Japan.
Email: may.sweet@qff.jp
Ph: (+81) 090-9699-4320, (+81) 050-7103-6063

**Abstract:** Neutron beam, being electrically neutral and highly penetrating, offers unique advantages for irradiation of biological species such as plants, seeds, and microorganisms. We comprehensively investigate the potential of neutron irradiation for inducing genetic mutations using simulations of J-PARC BL10, JRR-3 TNRF, and KUANS for spallation, reactor, and compact neutron sources. We analyze neutron flux, energy deposition rates, and Linear Energy Transfer (LET) distributions. KUANS demonstrated the highest dose rate of 17 Gy/h, significantly surpassing BL10, due to the large solid angle by the optimal sample placement. The findings highlight KUANS's suitability for efficient genetic mutations and neutron breeding, particularly for inducing targeted mutations in biological samples. The LET range of KUANS is concentrated in 20-70 keV/μm, which is potentially ideal for inducing specific genetic mutations. The importance of choosing neutron sources based on LET requirements to maximize mutation induction efficiency is emphasized. This research shows the potential of compact neutron sources like KUANS for effective biological irradiation and neutron breeding, offering a viable alternative to larger facilities. The neutron filters used in BL10 and TNRF effectively excluded low-energy neutrons with keeping the high LET component. The neutron capture reaction, $^{14}$N(n,p)$^{14}$C, was found to be the main dose under thermal neutron-dominated conditions.

## 1. Introduction

Genetic mutations are caused by alterations in the DNA sequence of a living organism. It occurs naturally and can be artificially induced to produce useful variants. Genetic mutations could be induced by various mechanisms: chemical mutagens, and ionizations by radiations.

One of the primary mechanisms is exposure to chemical mutagens, alkylating agents are used, such as ethyl methane sulfonate (EMS) treatment. EMS is advantageous in generating leaky alleles in forward genetics due to its ability to induce mutations [1][2], primarily causing G:C → A:T transition mutations during DNA replication. Although EMS treatment is beneficial for producing leaky alleles and enabling forward genetics inquiry, it is difficult to knock out numerous overlapping genes or modify promoter areas due to its narrow mutation spectrum, which mostly induced point mutations. Additionally, the high toxicity of EMS poses significant obstacles. Excessive treatment may damage or kill biological materials such as microorganisms, seeds and plants damage.

An alternative method for inducing mutations is irradiation treatment. Conversely, in chemical-free irradiation treatments, the use of radiation has gained popularity in irradiation breeding and the induction of genetic mutations in recent years. Irradiation treatment employs ionizing radiation sources such as UV [3], X-rays, gamma-rays, heavy-ion rays, and neutron beams to induce mutations in seeds, embryos, whole plants, or parts thereof, as well as microorganisms [4–8].

Radiation can cause significant damage to biological tissues, particularly to DNA. The mechanisms of DNA damage from radiation include single-strand breaks (SSBs) and double-strand breaks (DSBs). The dose of ionizing radiation applied to biomaterials range from 0.1 Gy to over 100 Gy, with cell death rates increasing from minimal (10-20%) at low dose to nearly complete cell inactivation at high doses. Radiation can be characterized by its Linear Energy Transfer (LET), which measures the energy deposition per unit length of radiation's path through tissue. Low-LET radiation causes sparse ionizations and predominantly SSBs, whereas high-LET radiation causes dense ionizations and more DSBs. High-LET radiation is more damaging and has a higher relative biological effectiveness (RBE) due to its ability to produce complete, clustered DNA damage, leading to higher cell death rates compared to low-LET radiation [9]. Therefore, the LET value is a critical factor influencing mutation rates. For instance, in *Arabidopsis thaliana* mutagenesis, it has been observed that a LET value of 30 keV/ μm is the most effective in inducing albino-mutants in the M2 generation [10], whereas the Minimal Ionization Energy deposition (MIP) due to the fast charged particle is just 0.2 keV/μm. Among radiation sources, X-rays and gamma rays have lower LET since the energy deposits are induced by electrons, while heavy ion and neutron beams have high LET due to the energy deposit by nuclei.

X-rays and gamma-rays are notable for their high material permeability, which allows for external irradiation. Treatment with low LET X-rays and gamma-rays has the advantage of enabling the irradiation of large quantities of biological materials. However, due to the sporadic nature of energy deposition by low LET rays, they tend to cause scattered DNA damage, resulting in a relatively narrow mutation spectrum. Consequently, the mutation induction rate with low LET rays is only increased to about 10 to 100 times the spontaneous mutation rate, which is about $10^{-5}$ to $10^{-6}$. X-ray and gamma-ray induces DNA damage relatively randomly and causes many types of mutations including based substitutions.

Heavy ion beams are expected to generate higher mutation frequencies due to their strong biological impacts. According to radiobiological considerations, it has been posited that the primary effect of heavy-ion beams is the initiation of DSBs [11,12]. In the 1950s to the 1990s, gamma-ray gained popularity for irradiation treatments [13,14]. In the 1990s to the 2010s, heavy-ion beams have garnered attention due to their high-LET, which cause localized ionization effects, and a wider range of mutations compared to X-rays or gamma-rays [15]. The LET along the ion's path results more localized and severe biological damage, causing complex DNA breaks and chromosomal rearrangements. This inclination towards heavy ions can be ascribed to their larger mass and charge, leading to increased energy deposition within the target material. The LET of heavy-ion beams for use in biological research ranges from 22.5 keV/μm to 4000 keV/μm in the RIKEN RI-beam factory[16]. However, heavy ions have low material permeability, their transmission distance is several tens of micrometers (μm) from the material surface. When heavy-ion beams are irradiated to a biological material, they must be placed on the irradiation surface without overlapping each other typically within 2 mm sample thickness. Thus, irradiation with heavy-ion beams requires precise placement of the objects in the path[17], and irradiating large numbers of objects is taking effort and time-consuming.

Neutrons have also been considered due to their high-LET outcomes with high biological effects and high mutation induction rate and irradiation effect [18,19]. Unlike heavy-ions, neutron beams provide significant benefits due to their higher penetration into tissues and materials. This allows for more uniform irradiation of biological materials even in solvents (culture medium or microorganism solutions, etc). In mutation induction, neutron beams are highly efficient, surpassing X-rays and gamma rays. Remarkably, they require lower doses (e.g., 1/20 of gamma ray doses) [20] for successful outcomes, proving advantageous in mutating microorganisms with largely unknown genomes. These neutrons can directly collide with atomic nuclei within DNA molecules, causing DSBs and other forms of damage. While X-rays and gamma-rays have significant penetration capacities, their LET values of 0.2 and 2.0 keV/μm for gamma-rays (cobalt-60) and X-rays (250 keV) [21], primarily causes SSBs and may not be effective at causing DSBs.

Fast neutron, with energies of 0.1 to 1 MeV typically induces results in mutagenesis of deletions ranging from a few bases to several million bases [22]. They exhibit a higher frequency of DSBs, consequently leading to a higher relative biological effectiveness [23] compared to gamma rays. Studies have revealed fast neutrons to be an exceptionally effective mutagen in plants and microorganisms [23,24]. These effects making neutron beams more effective in introducing desired genetic changes or traits in organisms.

In 1980's, neutron irradiation was generally performed in nuclear reactors, and more recently in the 21st century, fast neutrons generated by particle accelerators such as those at J-PARC (Japan Proton Accelerator Research Complex) have been actively used [25]. We have developed a bio-irradiation system at the particle accelerator and have been performing irradiation. Neutron energy distribution and flux at J-PARC BL10 are very well measured [26]. In this paper, we evaluated the dose and LET of neutrons on the sample using particle transport simulations (PHITS) [27].

Although the usefulness of J-PARC has already been demonstrated, the use of such a large facility is not only very costly, but also subject to restrictions such as access procedures that must be applied for months in advance, limited beam time, and sample take-out. In this situation, small neutron sources can be potential alternatives. Unlike nuclear reactors or spallation neutron sources, samples can be placed directly near the target, so they can have comparable performance despite their small size. Of particular note is the short cycle time due to the ability to irradiate at arbitrary times. Based on the success at J-PARC, the present study computes the irradiation dose for mutagenesis research using a compact neutron source, especially the Kyoto University Proton Accelerator Neutron Source (KUANS) as a model and compares the results with those of large facilities using simulations and compared with those of the large facility. The results showed that the performance of the compact neutron source can be equivalent or superior to that of the large facility.

In this paper, we present dose calculations for experimental setups at J-PARC BL10 and JRR-3 TNRF, which have experience with biosocial irradiation, and then compare the results with possible irradiation geometry with KUANS.

## 2. Method & Results
### 2.1 Irradiation setup at J-PARC and Simulation

Neutron irradiation experiments are being conducted at the MLF within J-PARC. A proton beam from the accelerator strikes a mercury target and generate neutrons through spallation reactions. These neutrons are then moderated and directed through beamlines. We used J-PARC BL10 (NOBORU)[25][22], which is a general-purpose beamline capable of fast neutron irradiation.

When neutrons interact with atomic nuclei, they induce some nuclear reactions: scattering, capture, or ion productions resulting a proton or an alpha. Radiation dose by neutrons is mainly caused by recoil protons produced by neutron scattering with hydrogen. Thus, high-energy neutrons above 1 keV were used to irradiate biological samples by cutting off slow neutrons. Figure1 illustrates the schematic diagram of the BL10 experiment and the associated simulation setup. In the left diagram, the illustration samples (yellow) indicate the irradiation cells, each containing biological materials. Fast neutron beams travel from the left side, retaining a 14 m distance to the irradiation cells. The neutron beam size is 7×7 cm$^2$, and the cylindrical shape samples size is 2.3 cm in length with a diameter of 55 cm. The neutron flux at the irradiation area is $3.2 \times 10^8$ n/cm$^2$/s in 1 MW equivalent [26].

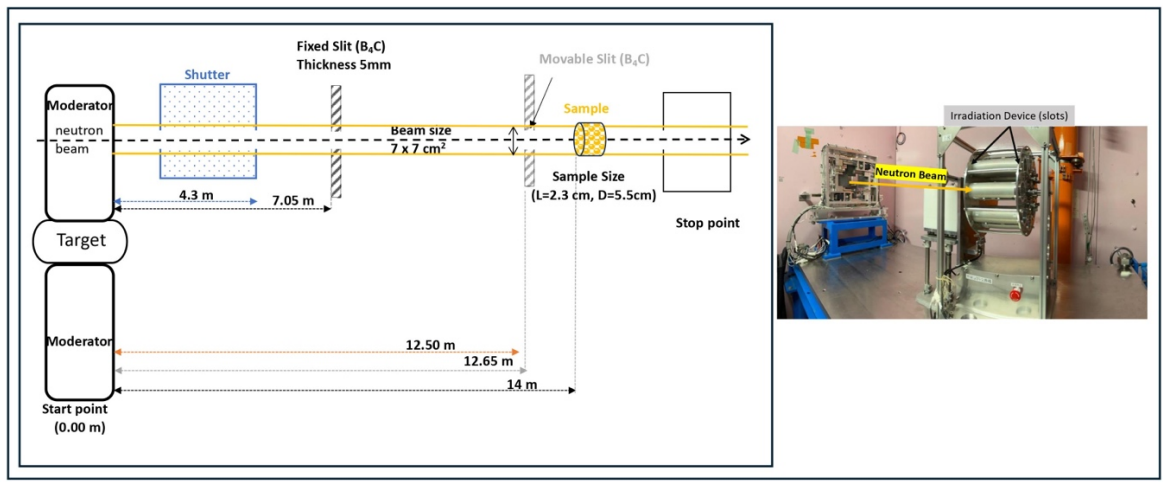

Figure1 Schematic side view of the neutron beam irradiation setup at J-PARC BL10 (left) and its photo (right). In the left diagram, the neutrons produce in the target travel to the beamline via the moderator. The sample shown in yellow indicates the irradiation cell containing biological materials. The $B_4C$ fixed slit and movable slit serve as an essential filter.

To evaluate the neutron irradiation process with the interactions between fast neutron beams and biological materials, we developed a simulation that replicated the irradiation setup at BL10 using PHITS, which is a general-purpose Monte Carlo particle transport simulation code that calculates neutrons, photons and electrons. The neutron flux with the energy spectrum was adopted from the reference [26], and the simulations were configured to reproduce it. The irradiation targets are plants, seeds, microorganisms, or microalgae. They are characterized by their chemical compositions, basically including carbon (C), hydrogen (H), nitrogen (N), oxygen (O) and others relevant elements. In this paper, we employed an irradiation sample with compositions C, H, N and O as 42.0%, 7.0%, 4.4% and 46.7% in weight% and weight density of 1.00 g/cm$^3$ to simulate typical bean (plant) seeds [28].

Because the slow neutrons do not contribute high LET but cause activation, we utilized a boron carbide filter ($B_4C$) of 5-mm thickness (Movable Slit in Figure. 1) to attenuate them [26]. We simulated the dose of neutrons with and without the $B_4C$ filter. The neutron energy spectrum calculated by PHITS are shown in Figure2(a). The neutron spectra are shown in a double-logarithmic graph, with the vertical axis described in lethargy units, which takes the natural logarithm as the unit. The spectrum without $B_4C$ shows a peak in cold neutron energy around 10 meV, then follows by a gradual increase up to ~1 MeV, where the energy that the generated spallation neutrons has. With $B_4C$, neutrons below 10 eV completely disappeared, but more than 70% neutrons survive in the vicinity of 1 MeV, which is useful for the irradiation. The total flux was attenuated to be $1.0 \times 10^8$ n/cm$^2$/s, mainly due to absence of the low energy neutrons.

Furthermore, we assessed the LET without and with $B_4C$ in Figure2(b) and Figure(c). The LET is classified as by electrons, protons and other nuclei. Scattering of neutrons with materials causes recoil nuclei, which become secondary charged particles with high LET which deposit energy in localized positions and causing significant biological damage.

In the spectrum of BL10 in Figure2(b), the energy deposition by recoil protons from neutron scattering with hydrogen nuclei is represented in the range 1-100 keV/μm in LET. The low LET component (~1 keV/μm) is due to the high-energy proton caused by scattering with high energy neutrons with ~100 MeV, while the high LET component is due to neutrons of ~1 MeV. The energy deposit by electrons is due to gamma rays from neutron absorption reactions, with a peak at the MIP region of 0.2 keV/μm. The LET corresponding to 100-1000 keV is due to recoil of nuclei such as C, N and O in the sample. In Figure2(c), additional contributions from heavy ions such as C, N, and O are highlighted. These heavy ions show significant energy deposition at higher LET values (100-1000 keV/μm). The total dose spectrum (black line) incorporates all particles, with peaks due to electron interactions at lower LET values and broader distributions from heavier ions and protons at higher LET values.

The total doses are obtained by integrating the LETs in Figure2(b) and Figure2(c). In case without $B_4C$ filter, the total energy deposit on the sample is $6.7 \times 10^6$ MeV/cm$^3$/s with 1MW beam power. Recoiled protons accounts for 85% of the total deposit. The effect of electrons, caused by neutron capture gammas, is 1.5% of the total; the total due to C, N and O recoils is 11%. A detailed discussion for the total doses is given in Section 4 with comparing other facilities.

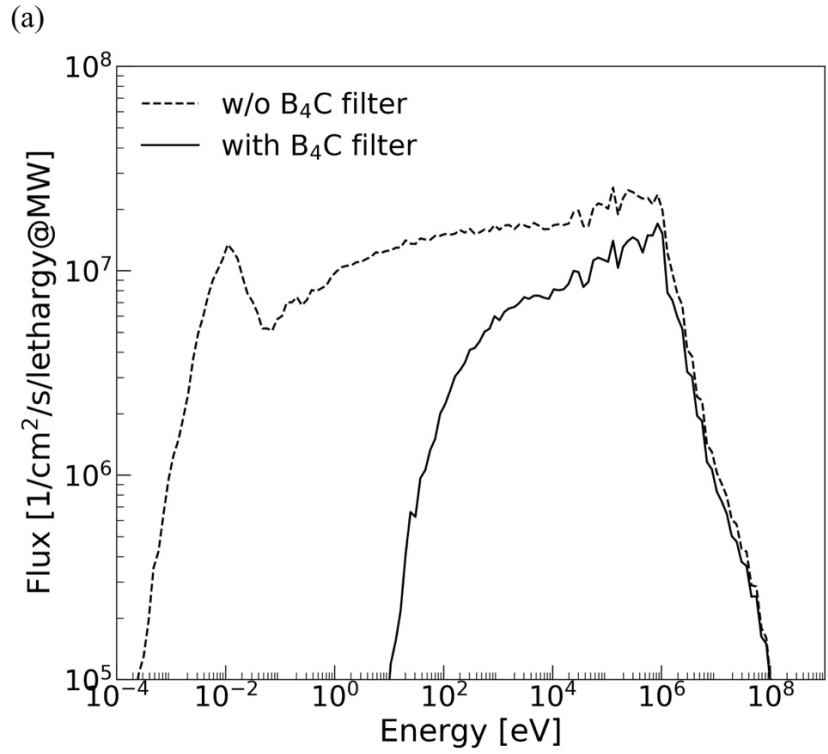

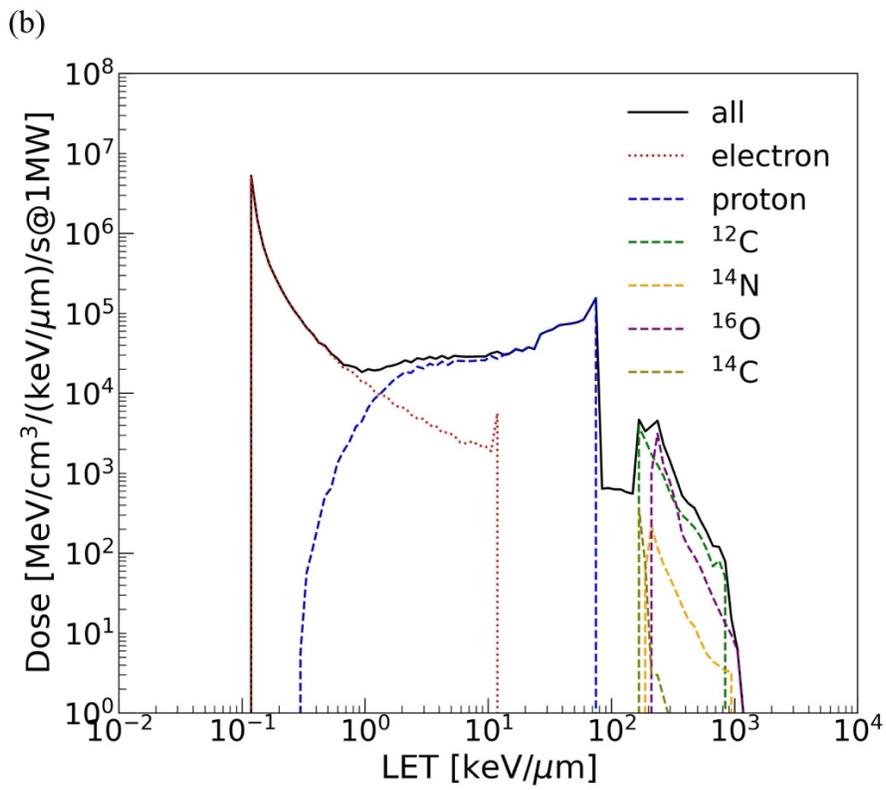

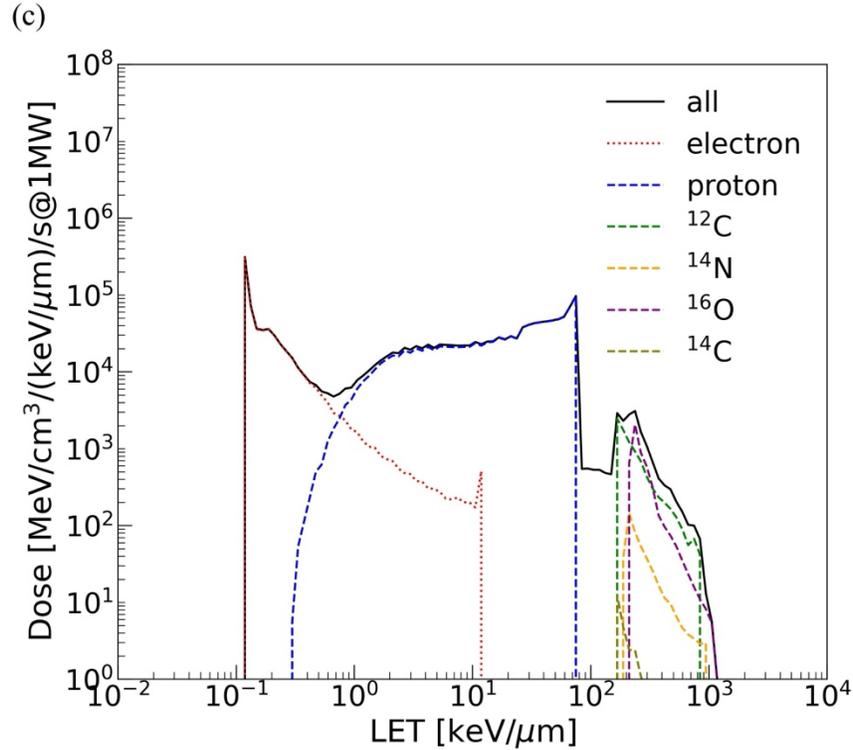

Figure2(a) Neutron flux as a function of energy with and without B$_4$C filter at J-PARC BL10. The neutron flux without the B$_4$C filter is depicted by the dashed line, whereas with the B$_4$C filter is represented by the solid. The LET distributions for different charged particles calculated with beam power of 1 MW in case without (b) and with (c) B$_4$C filter. The solid black line shows the total dose from all particles combined. The dotted red line represents the LET distribution of electrons, while the dashed blue line shows that of protons. Additionally, the green dashed line corresponds to $^{12}$C ions, the yellow dashed line to $^{14}$N ions, the purple dashed line to $^{16}$O ions, and the light green dashed line to $^{14}$C ions.

## 2.2 Irradiation setup at JRR-3 (TNRF)

Historically, most neutron irradiation has so far been carried out using nuclear reactors. We used TNRF porting JRR-3, In the JRR-3 experiments conducted at TNRF, which is a thermal neutron beam with a research reactor with heat power of 20 MW, was used to irradiate samples shown in Figure3. The neutron flux at TNRF port is at $1.0 \times 10^8$ n/cm$^2$/s with the beam size of W255×H305 mm [29]. This beam line produced the fast neutrons and gamma rays in the thermal neutron beam. For the irradiation at this beamline, a cadmium (Cd) plate of 0.5-mm thickness was used to attenuate the thermal neutrons to reduce activations of the sample materials.

We developed a simulation based on the BL10 case. With the same sample parameter, we used neutron flux and its energy spectrum at the beam area in reference [29] as input of the simulation. The neutron beam is from the reactor's core in water located 6.1 meters away with a beam size of $4 \times 4$ cm$^2$ and transported with widening along the beam path to $25 \times 20$ cm$^2$ at the sample position.

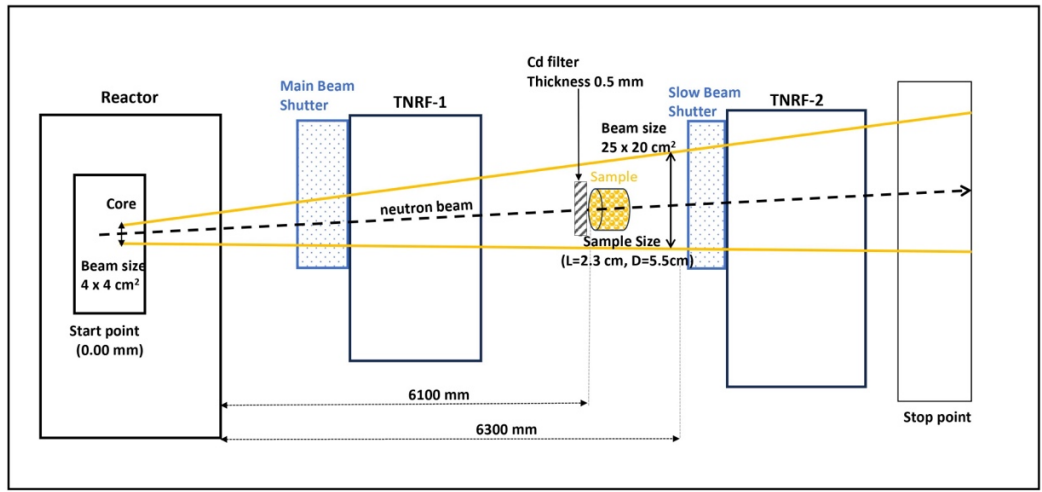

Figure3 Schematic of the neutron irradiation setup (side view) at JRR-3 (TNRF), which produces thermal neutron and fast neutron. The setup includes a Cd filter covering the irradiation samples to generate fast neutrons in advance. The sample shown in yellow indicates the irradiation cell containing biological materials.

(a)

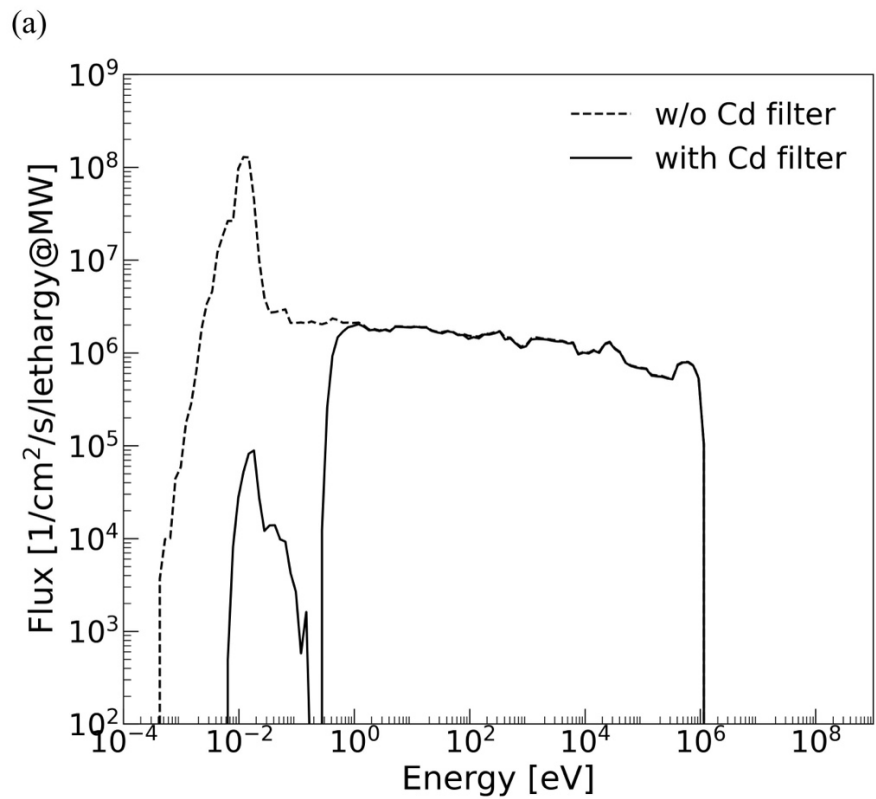

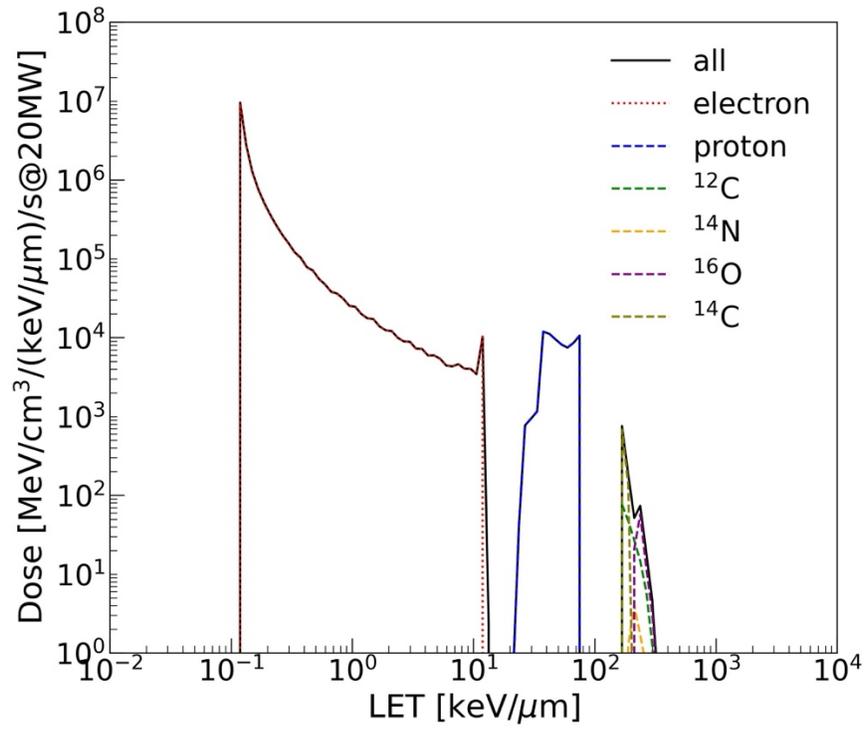

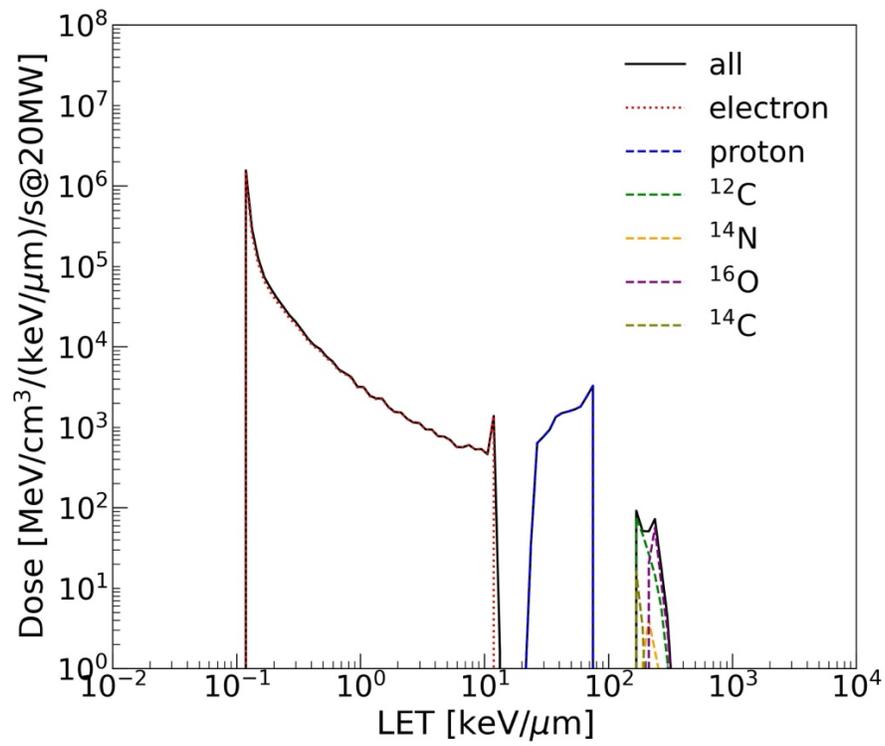

Figure4(a) Neutron flux as a function of energy with and without Cd filter at JRR-3, TNRF port. The dashed and solid lines represent the neutron flux without and with the Cd filter, respectively. The LET distributions for different charged particles calculated with reactor power of 20 MW in case without (b) and with (c) Cd filter. The solid black line indicates the total dose from all particles combined. The dotted red line represents the LET distribution of electrons, while the dashed blue line shows that of protons. Additionally, the green dashed line corresponds to $^{12}$C ions, the yellow dashed line to $^{14}$N ions, the purple dashed line to $^{16}$O ions, and the light green dashed line to $^{14}$C ions.

We calculated the neutron fluxes in case with and without the Cd filter. The calculated neutron energy spectra are shown in Figure4(a). The spectrum without the Cd filter, represented by a large peak around 10 meV indicating a high flux of thermal neutrons. The neutron flux gradually decreases up to about 1 MeV unlike BL10, covering a broad spectrum from thermal to high-energy neutrons. The use of the Cd filter of 0.5 mm alters the neutron flux spectrum by reducing thermal neutrons while preserving high-energy neutrons. This is due to the high absorption cross section of Cd, which effectively eliminate neutrons below 1 eV. The simulation result of total neutron flux with the Cd filter is $2.1 \times 10^7$ n/cm$^2$/s at 20 MW, which is 15% of that without Cd filter. The reduction is greater than that of BL10 (31%) because the main component at the TNRF is thermal neutrons.

The LET of JRR-3 without and with Cd filter are shown in Figure4(b) and Figure4(c), respectively. Protons contribute to the dose at higher LET as well as BL10 case but show a sharper peak in 10-100 keV/μm. It is because that the maximum energy of neutrons produced by a nuclear reactor is limited up to 2 MeV. In case with Cd filter, the total dose was $1.7 \times 10^5$ MeV/cm$^3$/s, of which 74% was due to protons and 17% to electrons. In case without the Cd filter, the total dose was $6.6 \times 10^5$ MeV/cm$^3$/s. Notably, the contribution of the $^{14}$N(n,p)$^{14}$C reaction accounted for 55% of the total dose. The effect of such ion emission reactions is more significant than recoil for thermal neutron irradiation.

**2.3 Simulation for a compact neutron source (KUANS)**

In recent years, many compact neutron sources have been constructed using small accelerators and Be or Li targets [30]. To explore the possibility of the neutron irradiation for the genetic mutation using a compact neutron source, we developed a simulation model for Kyoto University Accelerator-driven Neutron Source (KUANS) [31], which is early-type one and relatively have low power, as an example to estimate the irradiation dose.

KUANS consists of a proton linear particle accelerator (LINAC), a Be target, a polyethylene moderator, carbon reflectors, and radiation shields made by boron-doped polyethene and concrete. It has been developed for research and development of neutron devices, as well as for educational purposes. Here, we are considering the use of KUANS for biological irradiations. Neutrons in KUANS are generated through the $^9$Be (p, n) $^9$B reaction, using a pulsed 3.5MeV proton beam with and the maximum averaged current of 100 μA. The threshold energy of the neutron production reaction is 2057 keV and the maximum energy for the neutron is 1.3 MeV. The neutron production rate is expected as ~$10^{11}$ n/s [31], which is six orders of magnitude less than that of J-PARC (~$10^{17}$ n/s).

KUANS normally moderates the produced fast neutrons by a polyethylene moderator with a volume of $10 \times 10 \times 10$ cm$^3$ to obtain a thermal neutron beam. The moderated neutrons are transported perpendicular to the proton beam direction and used as a pulsed neutron beam[31]. Because unmoderated fast neutrons are useful for biological irradiation, we set up a computational geometry in our simulation by removing the polyethylene moderator and setting a sample just in front of the target where the generated neutrons hit it directly. The schematic diagram of KUANS for the simulation is illustrated in the Figure5. Fast neutrons are produced from Be target. The samples are located at 50 mm from the Be target to interact more fast neutrons. Because the Be target is covered with a graphite reflector, transmitted neutrons also have a certain probability of returning to the sample position. The neutron spectrum from the Be target was calculated by the reaction cross section [32] and proton energy loss in Be [33]. The neutron distribution was assumed to be isotropic.

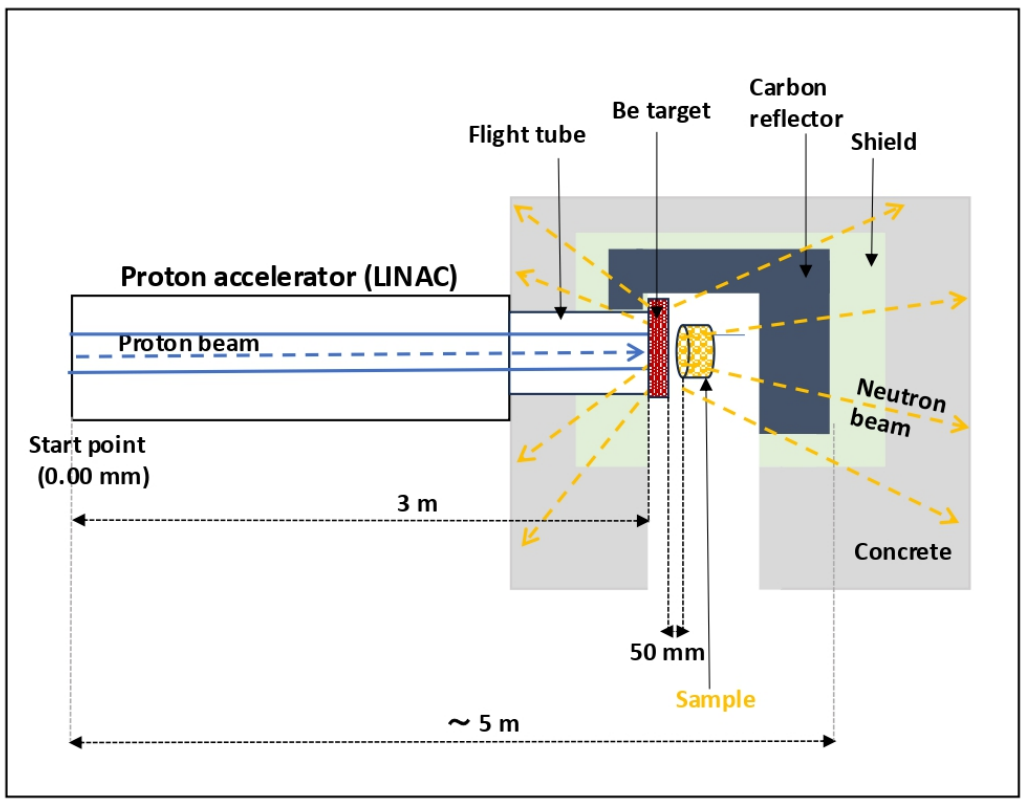

Figure5: Schematic diagram (side view) of KUANS. In the diagram, proton beams travel from the left side to the Be target (indicated red color), produced the neutron around. The sample shown in yellow indicates the irradiation cell containing biological materials.

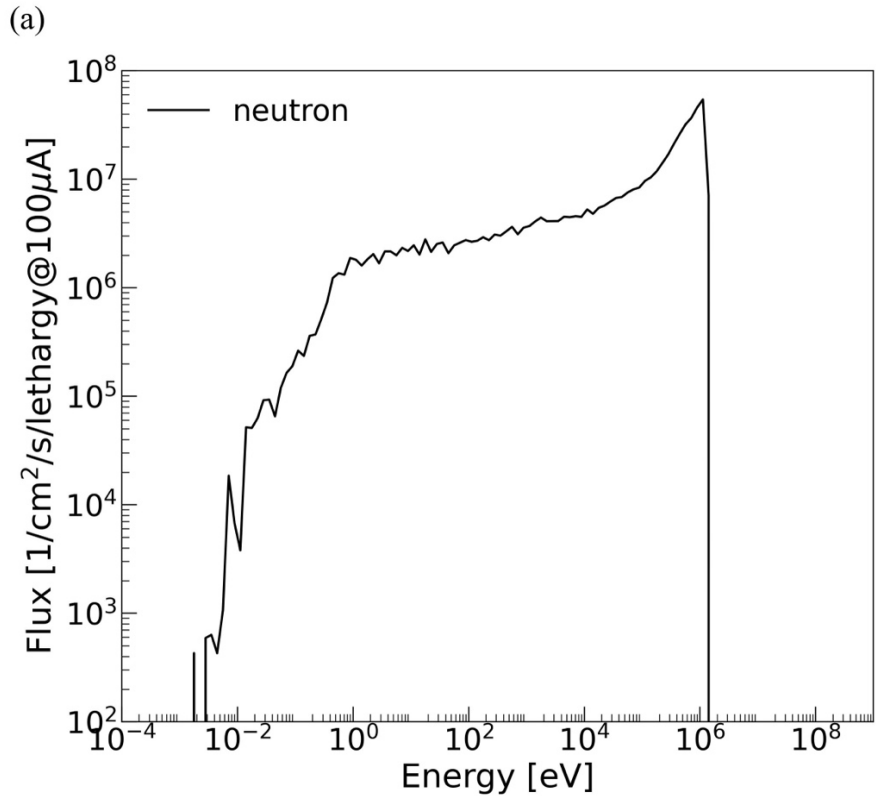

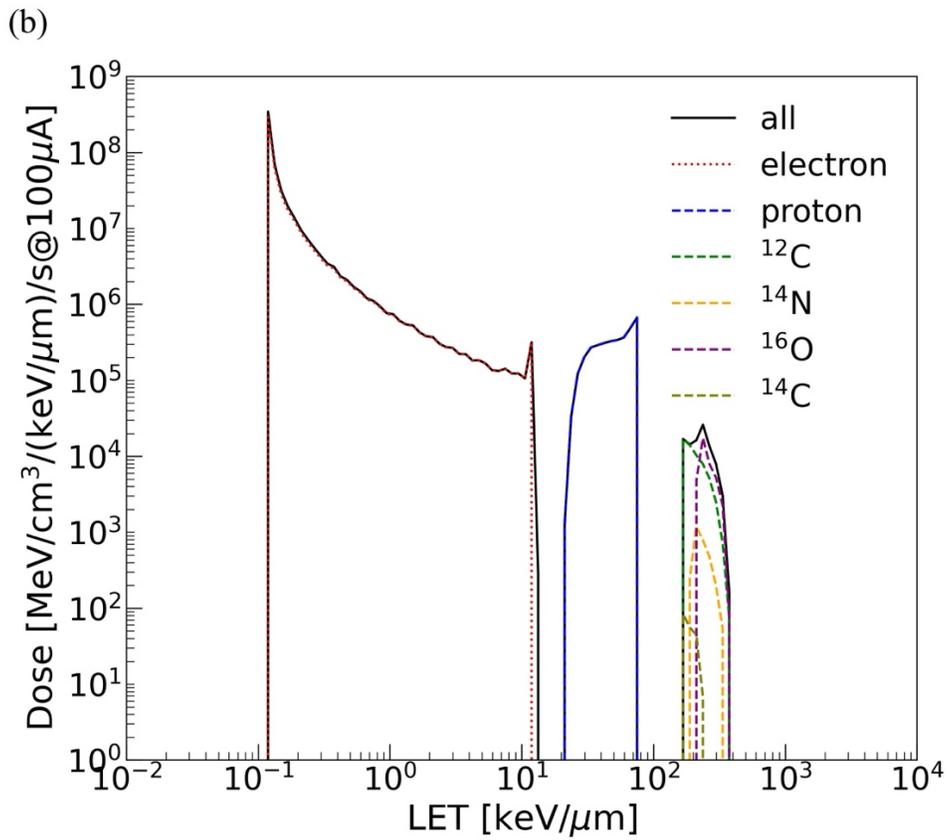

Figure6(a) Simulated neutron flux in the sample of the KUANS as a function of energy. (b) LET distributions for different charged particles. The solid black line represents the total dose from all particles, the dashed blue line is that of protons, and the dotted red line is for electrons. The green dashed line corresponds to $^{12}$C ions, the yellow dashed line to $^{14}$N ions, the purple dashed line to $^{16}$O ions, and the light green dashed line to $^{14}$C ions.

The neutron flux in the sample of the KUANS geometry is shown in Figure6(a). It has a high peak around 1 MeV, and thermal neutrons are vanishingly small due to the absence of moderators. In addition, the design does not use a moderator, so the thermal neutron flux is kept small even without using the thermal neutron filter like $B_4C$ or Cd. This is suitable for higher LET irradiation without causing activations. The neutron flux at KUANS is $4.6 \times 10^8$ n/cm²/s, which is 1.4 times higher than $3.2 \times 10^8$ n/cm²/s at J-PARC. This is because that the distance from the target to sample is very close (50 mm) in KUANS compared to 14 m in BL10, and larger solid angle are available. The LET by protons ranges in relatively narrower region of 20-70 keV/μm, as shown in Figure6(b), because the neutrons generated from the Be target are concentrated at 0.5-1.3 MeV.

## 3. Discussion

In the previous section, we have calculated the irradiation at J-PARC BL10, JRR-3 TNRF, and KUANS. The results are summarized in Table 1 for key parameters as neutron flux and energy deposition rates for each particle. In this section, we discuss the effectiveness for the biological irradiation at each site.

Neutron flux directly impacts the dose rate, and the irradiation time required to achieve specific biological effects. KUANS demonstrates a neutron flux of $4.6 \times 10^8$ n/cm²/s, which is 1.4 and 3.3 times higher than BL10 or TNRF, respectively. The most dominant factor in KUANS having the largest flux despite its less neutron generation rate is the distance from the target. By placing the sample at a distance comparable to the size of the target, a large solid angle of ~1 sr can be obtained, which is a significant advantage compared to using neutrons as a beam as in the case of large facilities (50 μsr for J-PARC BL10). In addition, the environment surrounded by reflectors but without moderators are a major advantage for use of the fast neutrons. Since LET of the slow neutrons is very low comparing to that of the fast neutrons, the effects of the moderators have little effects on the mutation.

The dose rate per unit time was the highest for KUANS, $3.0 \times 10^7$ MeV/cm³/s, despite the small neutron production rate. This is 4.5 times higher than the BL10 result without the neutron filter. In both cases, KUANS and BL10, the energy deposit is mostly due to recoil proton. Because the fraction of high-energy neutrons is less, so the total dose rate with filtered beam at TNRF was 1/31 of BL10 with the filter. Based on the density of the sample used of 1.00 g/cm³, the dose rate can be converted: 2.5, 0.08, and 17 Gy/h for BL10 (with $B_4C$), TNRF (with Cd), and KUANS, respectively. Since the typical dose for genetic modification is 40 Gy [23], KUANS can be completed the irradiation in 2.4 hours. This result highlights the potential for efficient irradiation with the compact neutron sources like KUANS. Low LET doses due to electrons are more likely to cause SSBs rather than DSBs and should be mitigated for efficient mutation. In the present calculation, the dose due to the electrons was suppressed to less than 30% of the total dose under all conditions.

Neutron filters used in BL10 or TNRF were found to be effective to exclude neutrons below eV while retaining MeV neutrons. Low-energy neutrons increase the dose by electrons induced from gamma rays, however, these only contribute to low LETs below 10 keV/μm. Therefore, it is expected that the contribution would be comparable to that of gamma irradiation. Thus, neutron filters allow irradiation only with high LET. With the $B_4C$ filter, the dose rate by electrons is reduced to 1/15 at BL10, while that by the proton is kept at 65%. The results of the TNRF without a filter showed that the energy deposition by the $^{14}N(n,p)^{14}C$ reaction with thermal neutrons was dominant. This reaction accounted for 54% of total dose, while the contribution by the recoil protons was 11%. The $^{14}N(n,p)^{14}C$ reaction emits proton of 583 keV and $^{14}C$ of 42 keV, giving LETs of 49-90 keV and 200 keV, respectively. Both have LETs high enough to induce DSBs in DNAs. This reaction may be responsible for the similar mutations observed with thermal neutron irradiation as with fast neutrons [34]. If ion-emitting nuclides with neutron capture, as $^6Li$ and $^{10}B$, can be implemented in the samples, high LETs are obtained even only with thermal neutrons.

Table 1: The simulation results for J-PARC, JRR-3 and KUANS

| Facility | Beam Power | Neutron flux [n/cm²/sec] | Total energy deposit [MeV/cm³/s] | Energy deposit by recoil protons [MeV/cm³/s] | Energy deposit by electrons [MeV/cm³/s] |
|---|---|---|---|---|---|
| J-PARC BL10 (w/o $B_4C$) | 1 MW | $3.2 \times 10^8$ | $6.7 \times 10^6$ | $5.7 \times 10^6$ | $1.0 \times 10^5$ |
| J-PARC BL10 (w/ $B_4C$) | 1 MW | $1.0 \times 10^8$ | $4.4 \times 10^6$ | $3.7 \times 10^6$ | $8.8 \times 10^3$ |
| JRR-3 TNRF (w/o Cd) | 20 MW (Thermal power) | $1.4 \times 10^8$ | $6.6 \times 10^5$ | $4.3 \times 10^5$ | $1.9 \times 10^5$ |
| JRR-3 TNRF (w/ Cd) | 20 MW (Thermal power) | $2.1 \times 10^7$ | $1.4 \times 10^5$ | $1.0 \times 10^5$ | $2.3 \times 10^4$ |
| KUANS | 3.5 MeV × 100 μA | $4.6 \times 10^8$ | $3.0 \times 10^7$ | $2.1 \times 10^7$ | $5.3 \times 10^6$ |

The biological effects are determined by genetic mutations induced by DSB of DNAs. Therefore, they are influenced by not only total dose but also LET. The optimum LET for mutation induction differs among biological species. For seeds of *Arabidopsis*, peak lethality occurs at LET values between 290-400 keV/μm [35,36], while mutation induction appears most effective at around 30 keV/μm [10,37,38] for inducing albino mutants. A LET of 50 keV/μm of $^{12}C^{6+}$ ions was ideal for achieving phenotype mutants in rice seeds [39]. In the case of *chrysanthemum*, reported that a LET of 76 keV/μm was most effective for inducing flower color mutants [40]. As discussed in the previous sections, neutron irradiation gives a different LET to heavy ions; neutron irradiation gives different LETs compared to heavy ions; typically, the LET of carbon (C) ions is ranging from 10 to 200 keV/μm and that of iron (Fe) ions is 100 to 500 keV/μm [10,36]. J-PARC and KUANS have different LET distributions. This is because the incident neutron energy is 0.5-1.3 MeV at KUANS, whereas J-PARC includes higher neutrons above 100 MeV. The LET in J-PARC is broadly distributed from 1 to 100 keV/μm by protons and 100-1000 keV/μm for $^{12}C$, $^{14}N$, and $^{16}O$, whereas in KUANS it is concentrated in the 20-70 keV range. Based on the results of this study, we can use different neutron sources depending on the required LET range to induce more efficient genetic mutations.

4. **Conclusions**

In this study, we developed a simulation that replicate real experimental setups to assess the potential of neutron irradiation in inducing genetic mutation for biological species. Simulations were conducted for J-PARC BL10 and JRR-3 TNRF port, which have already demonstrated genetic mutations, and the KUANS to investigate the potential of compact neutron sources, for the cases using a model of typical bean (plant) seeds as irradiation samples with PHITS code.

The main source of the dose was due to the recoil protons from fast neutron scattering with energy of around 1 MeV. Despite lower neutron production, KUANS achieved the highest dose rate of 17 Gy/h, which is 4.5 times of BL10. The low LET component by the electrons caused gamma rays are well suppressed in all conditions calculated. Since TNRF is customized for thermal neutrons and the fast neutrons are suppressed, the effect of recoil protons is small. However, in case TNRF without the Cd filter, the effect of $^{14}$N(n,p)$^{14}$C reaction was dominant, where it accounted 5 times larger than the recoil protons. This reaction may be responsible for the mutations with thermal neutron irradiation, and doping with $^6$Li or $^{10}$B, which absorb neutrons to produce ions as well, may enable genetic mutations even under thermal neutron irradiation alone like Boron Neutron Capture Therapy (BNCT).

The proton recoil, which is the main component of the neutron-induced dose, ranges from 1-100 keV/μm on LET distribution. The LET by protons for KUANS has a narrower peak in the 20-70 keV than that of BL10 because the neutron spectrum of KUANS are limited up to 1.3 MeV. We could potentially induce genetic mutations more efficiently by using different neutron sources depending on the required LET range.

**Declaration of competing interest:** The authors declare that they have no known competing financial interests or personal relationships that could have appeared to influence the work reported in this paper.

**Data availability:** Data will be made available on request.

**Acknowledgements:** The authors are deeply grateful to the accelerator operation staff and the experimental groups at J-PARC (BL10), JRR-3 (TNRF), and KUANS for their invaluable support during the experiment. They also extend their gratitude to Quantum Flowers & Foods Co., Ltd. for their financial support and project conceptualization. The authors acknowledge Takashi Ino and Katsuya Hirota for the collaboration research with the High Energy Accelerator Research Organization (KEK).